\documentstyle[epsf,graphicx]{mn}

\def\nh{N$_{\rm H}$} 
\def\lunits{erg~s$^{-1}$}
\def\funits{erg~cm$^{-2}$~s$^{-1}$}
\def\cunits{cm$^{-2}$}
\def\xmm{{\it XMM-Newton~}}
\def\chandra{{\it Chandra~}}

\def\asca{{\it ASCA~}}

\begin{document}

\title[SHEEP] 
{XMM-Newton and Chandra observations 
 of SHEEP sources}

\author[Georgantopoulos et al.] 
{ {\Large  I. Georgantopoulos$^1$,   K. Nandra$^2$, 
M. Brotherton$^{3,4}$,  A. Georgakakis$^{1,2}$,  
I. E. Papadakis$^5$,  P. O'Neill$^2$}  \\
$^1$Institute of Astronomy \& Astrophysics, 
National Observatory of Athens, I. Metaxa \& V. Pavlou, Athens, 15236, Greece \\
$^2$Astrophysics Group, Imperial College London, Blackett Laboratory,
  Prince Consort Rd, SW7 2AW  \\
$^3$ Department of Physics \& Astronomy, University of Wyoming, Laramie, WY82071, USA \\
$^4$ Kitt Peak National Observatory, 950 N. Cherry Avenue, Tucson, AZ 85719, USA \\
$^5$ Physics Department, University of Crete, 71003, Heraklion, Greece \\}

\maketitle

\begin{abstract}
We present \chandra and \xmm observations of 12 bright  ($\rm f(2-10 keV)
 > 10^{-13}$ \funits) sources from the \asca SHEEP (Search for the High Energy
 Extragalactic  Population) survey.  Most of these have been either not
 observed or not detected previously  with the  {\it ROSAT} mission and therefore they
 constitute a sample biased towards   hard sources.  The \chandra observations are
 important in  locating with  accuracy the optical counterpart of the X-ray sources.  
 Optical spectroscopic observations show that  our sample is associated with both
 narrow-line (NL) (six objects),   and Broad-Line (BL) AGN (five objects) with one
 source  remaining unidentified.  Our sources cover the redshift range  0.04 to 1.29
 spanning luminosities from $10^{42}$ to $10^{45}$ \lunits (2-10 keV). The NL sources
 have preferentially  lower redshift (and luminosity) compared with the BL ones. This
 can be most easily explained in a model where the  NL AGN are intrinsically less
 luminous than the BL ones in line with the results of Steffen et al. 
 The X-ray spectral fittings show a roughly equal number  of
 obscured ($N_H>10^{22}$ \cunits) and unobscured  ($N_H<10^{22}$ \cunits) sources.
 There is a clear tendency for obscured sources to  be associated with NL  AGN and 
 unobscured sources with BL ones.
  However, there is a  marked exception with the 
 highest obscuring column observed at a BL AGN  at a redshift of z=0.5. 
  \end{abstract}


\section{Introduction}
 
 \chandra surveys resolved 80 per cent of the X-ray background  at hard energies
 (i.e. 2-10 keV) Brandt et al. (2002),  Giacconi et al. (2002), Alexander et al.
 (2003), shedding ample light on its origin.  Optical identifications of the
 X-ray sources are showing  that these are mainly AGN. The largest fraction of
 these are absorbed in X-rays, presenting an average spectrum with a slope of
 $\Gamma=1.4$ (Tozzi et al. 2001) similar to the slope of the X-ray background. 
 Large area bright surveys performed  with XMM-Newton in the 2-10 keV band are
 complementary   as they help us to explore the bright, nearby counterparts  of
 the sources detected in the deep Chandra fields. One remarkable finding from
 these bright surveys  is the scarcity of obscured AGN ($\rm N_H>10^{22}$
 \cunits)  at bright fluxes (Piconcelli et al. 2002, Georgantopoulos et al.
 2004, Perola et al. 2004). This comes in contrast to the predictions of the 
 population  synthesis models (e.g. Comastri et al. 1995, Gilli et al. 2001).  

 Selection in the hardest band allowed by imaging surveys  (5-10 keV)
 facilitates the selection of the most obscured  candidates, as this bandpass is
 relatively unbiased to absorption.  Indeed columns as high as a few times
 $10^{23}$ \cunits ~ do not   practically affect X-ray energies above 5 keV.   
 The HELLAS survey (Fiore et al. 1999)  pioneered these studies using the MECS
 detector on board {\it BeppoSAX}.  The HELLAS survey covered  an area of 85 $\rm
 deg^2$ detecting about 150 sources  down to a flux limit  of $5\times 10^{-14}$
 \funits in the 5-10 keV band. 

 The SHEEP survey further explored this energy bandpass with {\it ASCA} GIS. It
 comprises of 69 objects serendipitously  detected in a non-contiguous area of
 39 $\rm deg^2$ down to a flux limit of  $\sim10^{-13}$\funits in the 5-10 keV
 band (Nandra et al. 2003).  35 of these objects were detected by {\it ROSAT},
 in either pointed observations or in the All Sky Survey, 
 with 13 of them  having secure optical counterparts in optical  catalogues. 
 For the remaining 34 sources,  which were either not observed or not detected 
 by {\it ROSAT}, it is very difficult  to locate the optical counterpart as the
 {\it ASCA}  error box is large (typically 60 arcsec rms). We have thus obtained
 \chandra snapshot observations  in order to  extract reliable positions ($<$1
 arcsec) of the X-ray source that will enable us to search for their optical 
 counterparts.  Preliminary results from  these  observations are presented in
 Nandra et al.~(2004).

 Here, we report our results from X-ray and optical observations of 12 out of the
 34 \chandra sources which have a $2-10$ keV flux larger than $10^{-13}$ 
 \funits (estimated assuming a Photon index $\Gamma=1.9$). 
 We have obtained optical spectroscopy for 11 of these objects. Furthermore, 
 eight of these have been also observed
 by \xmm, providing  good quality spectra. Consequently, for these bright
 sources, we can derive their X-ray spectral parameters (i.e. \nh\ and
 $\Gamma$) using proper spectral fits with good photon statistics (instead of
 hardness ratios only). Moreover,  Nandra et al.~(2004) noticed that a
 fraction of the \chandra sources are much fainter  (in some cases more than
 an order  of magnitude) than the original {\it ASCA} source. Although
 variability could play some role,  there is clearly some ambiguity about the
 reality  of these sources. Watanabe et al. (2002) who performed  \chandra
 pointings of hard {\it ASCA} sources also report the same problem. By choosing
 to study initially only the bright sources, we minimize the possibility that
 a source is spurious. The optical counterparts for three of the sources in
 the present SHEEP subsample have already been presented in Nandra et al.
 (2004).We include them in the present work as they satisfy the flux limit we
 have set, and we also present the results from \xmm observations of them.

\section{Data Acquisition}

\subsection{The Chandra observations} 
 
Our sources were observed using  the Advanced CCD Imaging Spectrometer,
ACIS-S, on board the {\it Chandra} observatory (Weisskopf 1997). We use the
event file provided  by the standard pipeline processing.  Only Grade 0,2,3,4
and 6 events are used in the analysis. Charge Transfer Inefficiency (CTI)
problems do not  affect our observations as S3 is a back-illuminated chip.
Each CCD chip subtends an 8.3 arcmin square on the sky while the pixel size is
$0.5''$.  The spatial resolution on-axis is $0.5''$ FWHM.  The ACIS-S spectral
resolution is  $\sim$100 eV (FWHM) at 1.5 keV. We  estimate that even in the
brightest case (AXJ1531.9+2420) the pile-up fraction is less than 12 per cent.
Images, spectra, ancillary files, response matrices,  and  light curves have
been created   using the {\sl CIAO v2.2} software. We use a $2''$ radius
extraction region in order to produce  both the spectral files and the
light curves. We take into account the degradation of the ACIS quantum 
efficiency in low energies, due to molecular contamination,  by using the {\sl
ACISABS} model in the spectral 
fitting\footnote{http://asc.harvard.edu/cal/Acis/Cal\_prods/qeDeg}. The
observation details are given in Table \ref{chandra-log}.

\begin{table*}
\footnotesize
\begin{center}
\begin{tabular}{cccccc}
name & RA$^1$ & Dec$^1$ & Sequence$^2$ & Exposure$^3$ & Date \\
\hline 
AXJ0140.1+0628 & 01 40 10.1 & +06 28 27 &  900149 & 5.8  & 2002-02-07  \\
AXJ0144.9--0345 & 01 44 55.4 & -03 45 23 & 900150 & 4.6  & 2002-02-07  \\ 
AXJ0335.2--1505 & 03 33 18.1 & -15 06 17 & 900152 & 4.9  & 2002-02-19 \\
AXJ0440.0--4534 & 04 40 01.9 & -45 34 09 & 900154 & 4.7  & 2002-06-22 \\
AXJ0836.2+5538 & 08 36 22.8 & +55 38 41 &  900155 & 4.4  & 2001-11-10\\
AXJ1035.1+3938 & 10 35 15.6 & +39 39 09 &  900158 & 5.0  & 2002-03-26  \\
AXJ1230.8+1433 & 12 30 51.8 & +14 33 23 &  900162 & 5.2  & 2002-07-21 \\
AXJ1406.1+2233 & 14 06 07.0 & +22 33 34 &  900167 & 4.9  & 2002-08-05 \\   
AXJ1511.7+0758 & 15 11 49.9 & +07 59 19 &  900172 & 5.1  & 2002-07-22 \\
AXJ1531.8+2414 & 15 31 52.3 & +24 14 30 &  900174 & 5.5  & 2002-10-08 \\
AXJ1531.9+2420 & 15 31 59.0 & +24 20 47 &  900175 & 5.1  & 2002-09-25 \\
AXJ1532.5+2415 & 15 32 33.1 & +24 15 26 &  900177 & 4.9  & 2002-07-02  \\
\hline 
\multicolumn{6}{l}{$^1$ X-ray Equatorial Coordinates J2000; $^2$ Chandra Sequence number;
 $^3$ Exposure in ksec;} \\ 
\end{tabular}
\end{center}
\caption{The list of the 12 SHEEP targets studied in this work, and the details of
the \chandra observations. }
\label{chandra-log} 
\normalsize
\end{table*}

\subsection{The XMM-Newton Observations} 

The X-ray data have been obtained with the EPIC (European Photon Imaging
Camera; Str\"{u}der et al. 2001 and Turner et al. 2001) cameras on board the
XMM-{\it Newton} operating in full frame mode. 
Four sources have been observed by us while another four 
 have been recovered from the XMM-Newton archive. 
 The observational details are
given in Table \ref{xmm-log}. The data have been analyzed using  the Science
Analysis Software ({\sc sas 5.3}). Event files for both the PN and the MOS
detectors   have been produced using the {\sc epchain} and {\sc emchain} tasks
of {\sc sas} respectively. The event files were screened for high particle 
background periods.  In our analysis we have dealt only with events
corresponding to patterns 0-4 for the PN and 0-12 for the MOS instruments. 

The source spectra are extracted using an 18\,arcsec radius circle. This area
includes at least 70 per cent of the X-ray source photons at off-axis angles
less than 10\,arcmin. 
The background spectral files are extracted from every image  independently,
using regions free from sources with a total area  about 10 times larger than
the source area. The response matrices and the auxiliary files  are produced
using the {\sc sas} tasks {\sc rmfgen} and {\sc arfgen} respectively. 
AXJ1035.1+3938 has just been detected by \xmm and thus we cannot derive 
 a spectrum as there are not sufficient counts. The flux of this 
 source is $F(2-10 keV) =6.8\times10^{-14}$ \funits. 

\begin{table*}
\footnotesize
\begin{center}
\begin{tabular}{cc cc cc c}
\hline
 Name & Obs-ID & FILTER & PN$^1$  & MOS$^1$  &  Field name & Date \\
\hline
AXJ0140.1+0628$^{+}$ & 0110890901 & MED & 23.0 & 26.9 & PHL1092 & 2003-01-18 \\
AXJ0836.2+5538       & 0143653901 & MED & 7.1 & 9.5 & AXJ0836.2+5538 & 2003-10-09 \\
AXJ1035.1+3938 $^{+}$ & 0109070101 & THIN& 12.5 & 15.1 & RE1034+396 & 2002-05-01 \\
AXJ1230.8+1433 $^{+}$ & 0106060601 & THIN & 8.9 &      & VIRGO6 & 2002-07-08 \\
AXJ1406.1+2233 $^{+}$ & 0051760201 & THIN & 14.7 & 17.3 & PG1404+226 & 2001-06-18 \\
AXJ1531.8+2414 & 0143650101 & THIN & 4.1 & 5.9 & AXJ1531.9+2420 & 2003-08-03 \\
AXJ1531.9+2420 & 0143650101 & THIN & 4.1 & 5.9 & AXJ1531.9+2420 & 2003-08-03 \\
AXJ1532.5+2415 & 0143650101 & THIN & 4.1 & 5.9 & AXJ1531.9+2420 & 2003-08-03 \\
\hline
\multicolumn{7}{l}{$^1$ Exposure Time in ksec} \\
\multicolumn{7}{l}{$^+$ Archival Observation} \\
\end{tabular}
\end{center}
\caption{The XMM-{\it Newton} observations}
\label{xmm-log}
\normalsize
\end{table*}

\subsection{The optical observations}
 
Optical imaging of our targets in the Johnson B and R band has been performed
with the Skinakas 1.3-m/f7.7 Ritchey-Cretien telescope of the University of
Crete, except  from AXJ0335.2--1505 and AXJ0440.0--4534 which were observed
with the ANDICAM detector in the SMARTS 1.3 m telescope at CTIO, Cerro-Tololo 
 Interamerican Observatory. All images
were reduced in the same way, as explained by Nandra et al. (2004). Typically,
all observations were carried out through standard Johnson {\it B} and Cousins
{\it R} filters. Exposure time was 20 min for both filters. In all cases,
observations were carried out under photometric conditions, with seeing being
between $\sim 1-1.5-$arcsec, and an optical source was detected within 1
arcsec of the \chandra source. The first, second, fifth, and the last two
sources in Table \ref{opt-log} appear point-like, while the other seven
sources are extended, with dimensions ranging between $\sim 10$ up to $\sim
30$ arcsec (in the case of AXJ0440.0--4534). The integrated $R-$band
magnitudes, and $B-R$ colours of all sources are listed in Table
\ref{opt-log}. 

\begin{figure*}
\includegraphics[width=17cm,height=15cm]{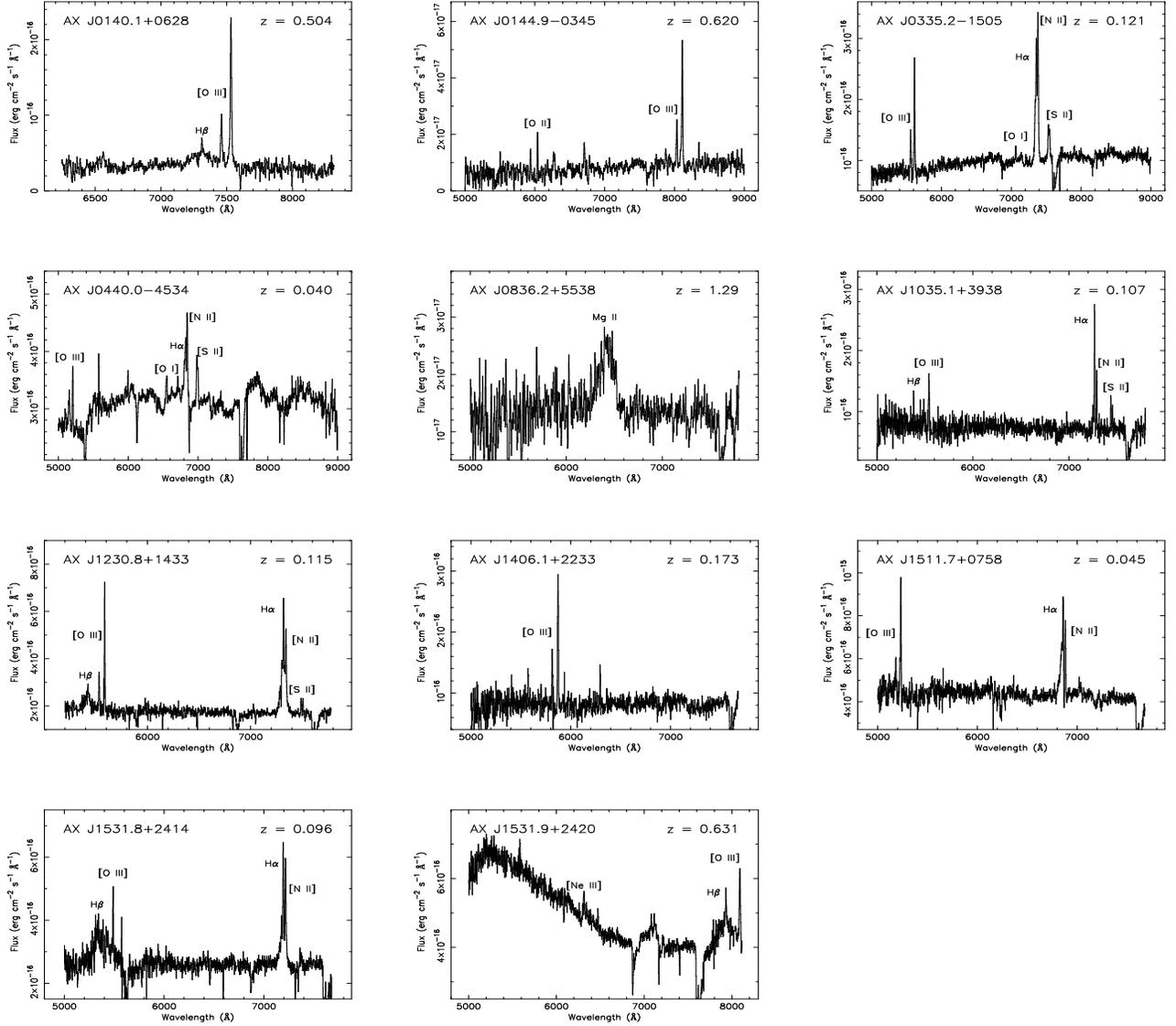}
\caption{The optical spectra}
\label{optical}
\end{figure*}

 Optical  spectroscopy  of eight sources in the Northern Hemisphere  has  been
obtained with the  Ritchey-Cretien Focus Spectrograph on the 4-m Mayall
telescope  at the Kitt-Peak National Observatory.  The spectral coverage was
from 5000 to 8000 \AA,  with a resolution of about 300 $\rm km~s^{-1}$.  The
optical spectra for three of these  were presented in Nandra et al. (2004). 
Three more sources have been obtained with the Cassegrain  spectrograph  at the
CTIO 4-m V. N. Blanco telescope on 9 Dec 2003.  We used the R-C spectrograph
with the G181 grating (316 l/mm), blazed at 7500 Ang, and the order-blocking
filter GG-495. There is only one camera-CCD combination that can be used, which
is the Loral 3K CCD with the Blue Air Schmidt camera. The slit-width was 1.5"
and the spectral resolution was about 250 km/s. The spectral coverage is
5000-9700 Ang. The optical spectra show that five sources present Broad Lines
 ($\rm FWHM > 1000 km ~s^{-1}$) and therefore are clearly associated with AGN. 
 Another six present only Narrow Lines. It is difficult to classify these 
 as in some cases the $H\alpha$ or the $H\beta$ lines happen to 
 lie outside the spectral window. However, their X-ray properties
 i.e.  high X-ray luminosities ($\rm L_x>10^{42}$ \lunits) combined \
 with the hard spectra, suggest that most are AGN.  
The redshifts and optical classification (BL or NL)  are given
in Table \ref{opt-log}. The spectra are presented in Fig. \ref{optical}.

\begin{table*}
\begin{center}
\begin{tabular}{cccccc}
\hline
name & ID$^1$ & z$^2$ & R$^3$ & B-R$^3$ & Extended$^4$\\
\hline 
AXJ0140.1+0628  & BL & 0.504   & 18.9 & 0.9 & N \\
AXJ0144.9-0345  & NL & 0.620 & 20.1 &2.4 &    N\\ 
AXJ0335.2-1505  & NL & 0.121  & 17.5 & 1.5 &  Y \\
AXJ0440.0-4534  & NL &  0.040 & 13.7 & 1.7 &  Y \\
AXJ0836.2+5538  & BL & 1.290  & 20.2 & 1.2 & N \\
AXJ1035.1+3938  & NL & 0.107  & 18.1 & 1.4 & Y \\
AXJ1230.8+1433  & BL & 0.115 & 16.7 & 1.4 & Y \\
AXJ1406.1+2233  & NL & 0.173  & 17.7 & 2.0 & Y \\   
AXJ1511.7+0758  & NL & 0.045 & 15.3 & 1.2 & Y \\
AXJ1531.8+2414  & BL & 0.096   & 16.9 & 1.7 & Y \\
AXJ1531.9+2420  & BL & 0.631  & 17.2 & 0.5 & N \\
AXJ1532.5+2415  & -  & -     & 19.6 & 2.3 & N \\
\hline 
\multicolumn{5}{l}{$^1$  Spectral identification; $^2$ Redshift} \\
\multicolumn{5}{l}{$^3$ Johnson R magnitudes and B-R colours}   \\
\multicolumn{5}{l}{$^4$ Optical Extension}   \\
\end{tabular}
\end{center}
\caption{Optical Properties}
\label{opt-log} 
\normalsize
\end{table*} 

\section{The Data Analysis} 

We investigate the X-ray properties of our sources by performing spectral
fittings with the {\sl XSPEC} software package  v11.2.   The poor count
statistics do not allow the use of the standard $\chi^2$ analysis in all
cases. Instead, we use in some cases  the C-statistic technique (Cash 1979),
which is proper for  fitting spectra with limited number of counts. Note that
this method can be used to estimate parameter values and confidence regions
but does not provide a goodness-of-fit (Arnaud 1996). 

We fit the \chandra and \xmm data separately for each source in the  0.3-8 keV
band. In the \xmm case we fit  simultaneously the PN and the MOS data for each
source.  We use a power-law model with two absorption components  (wa*zwa*po
in XSPEC notation) to fit the data. The column density for the first
absorption component is fixed to the Galactic value  (the values used for each
field were taken from Dickey \& Lockman (1990) and are listed in Table
\ref{chandrafits}) while the $N_{H}$ for the second component (representing 
the intrinsic absorption) is left as a free parameter during the model fitting
process.   

In the case  of the \chandra data where we have limited photon statistics, we
fix the photon index to 1.9 and allow both the $\rm N_H$ and the  normalization
to vary.  In the case of \xmm sources with adequate photon statistics  we leave
both the column density and the photon index free  to vary. The results for
\chandra and \xmm  are presented in Tables \ref{chandrafits} and \ref{xmmfits} 
 respectively. All quoted errors  correspond to the 90 per
cent confidence level. The $N_H$ values listed in these tables correspond to
the column density of the intrinsic component. The luminosities have been 
estimated assuming $\rm H_o=70, \Omega_m=0.3 \Omega_\Lambda=0.7$. 
The X-ray spectra are presented in Fig. \ref{xspec}. 
 We see that in some sources that there are residuals which may 
 denote the presence of emission lines. However, in most cases  
 these are hardly statistically significant. 
 The stronger case for a line is in AXJ1531.9+2420, the brightest 
 source in our sample: an FeK line at 
 6.4 keV (rest-frame) is statistically significant at just above the 
 95 per cent confidence level.  Again a MgXI line at 1.35 keV (rest-frame) 
 is statistically significant at the 93 per cent confidence level.

 In the case of those sources where we have many photons to perform a model
 fitting with the standard $\chi^2$ analysis, we find that the model fits well
 the data except from the \chandra fits of AXJ1511.7+0758 and
 AXJ1531.8+2414. The second source has also been observed by \xmm, and the
 model provides a good fit to the data.  
 The best fitting spectral slope ($\Gamma=2.4$) is
 steeper than the $\Gamma=1.9$ value that we used for the model fitting of the
 \chandra spectrum. When we leave the photon index free in the \chandra 
 spectral fit we obtain $\Gamma=2.40^{+0.24}_{-0.13}$ with $\chi^2=38.3/23$
 in excellent agreement with \xmm. The other source,
 AXJ1511.7+0758, is the less luminous in our sample. Its optical counterpart
 has a NL spectrum. The relatively low luminosity combined with the 
 steep X-ray spectrum raises the possibility that a large fraction 
 of the X-ray emission comes from star-forming processes.
 Thus we have added a Raymond-Smith component to the X-ray spectrum. 
 The addition of this component significantly improves the fit ($\chi^2\approx 36.9/20$
 for $\rm kT\approx 0.2$ keV). Leaving free the photon index as well  
 results in a further improvement ($\Delta \chi^2 \approx 5.3$).
 The best fit photon index becomes $\Gamma=1.46^{+0.22}_{-0.18}$.

 The absorbed flux estimates from the \chandra and \xmm spectra are in
 agreement, except from the case of AXJ1230.8+1433, AXJ1406.1+2233, and
 AXJ1532.5+2415, which show a flux variation of the order of $\sim$ 0.6, 2.5 
 and 3, respectively, between the \chandra and \xmm observations. The two
 observations are separated by about 0.5, 10 and 13 months, respectively. 
 The last object is still spectroscopically unidentified  and hence 
 we cannot estimate its luminosity. On the other hand, the other two 
 objects are of moderate
 luminosity (3 and $9\times 10^{43}$ \lunits respectively). Similar amplitude
 variations are observed in nearby BL objects on similar time scales. 
 There is also a reasonable agreement, within the errors, between  the \chandra  and
 the \xmm  $N_H$ best fitting values.

 Most objects in the current sample show some amount of intrinsic absorption.
 Among the 11 objects with optical identifications, six are ``absorbed" systems
 (i.e. they have a column density of $>10^{22}$ \cunits).
 Nevertheless, five ``soft", i.e. $N_H<10^{22}$ \cunits, sources can  still be
 found in our sample. Sources AXJ1511.7+0758, AXJ1531.8+2414 and
 AXJ1531.9+2420  have not been  observed by {\it ROSAT}. As they are soft
 bright sources, they  would probably have been detected. Source AXJ1035.1+3938
 was lying within the field-of-view  (4.5 arcmin off-axis)  of a {\it ROSAT}
 HRI pointing (1.7 ksec) but was not detected.  
 Finally, source AXJ1230.8+1433 has been detected by {\it ROSAT} HRI. 

 In general, optically classified NL objects show more obscuration when
 compared to the BL objects. The median column density  of the former
 class of objects in our sample  is $\sim 2\times 10^{22}$ \cunits, compared 
 to $\sim 7\times 10^{20}$ \cunits for the BL AGN. In fact, using 
 a Kolmogorov-Smirnov test we find that the column density distribution of the 
 NL and BL samples are different at the 87 per cent significance level.
  This is not a highly significant result (because of 
 the small number of sources in our sample) but is nevertheless indicative
 that the trend we observe, i.e. $N_{H,NL}> N_{H,BL}$, may correspond to a real
 characteristic of the SHEEP sources. However, there are two clear exceptions
 to the general trend. The object which presents the highest X-ray obscuration
 is  a BL AGN (AXJ0140.1+0628 at z=0.5).  Reversely, one of the objects with the 
 lowest obscuration is a NL source at redshift of 0.045 (AXJ1511.7+0758).   

\begin{figure}
\rotatebox{270}{\includegraphics[width=6.5cm]{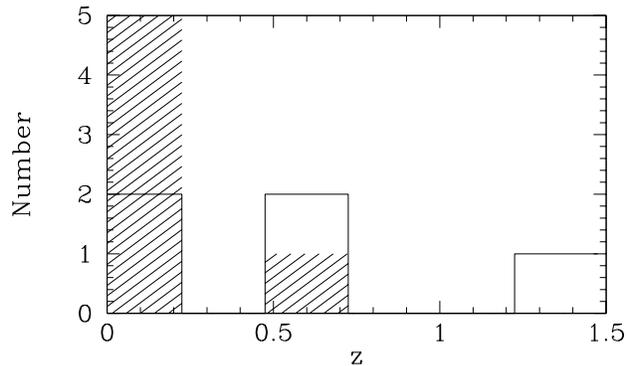}}
\caption
 {The redshift distribution of the BL (NL) AGN 
 denoted with open (hatched) histogram. 
 }
\label{z}
\end{figure}

 Finally, there is a separation in redshift between BL and NL populations
 in our sample.  The NL AGN are preferentially found at low redshifts  (median
 redshift z=0.11) while the BL AGN  are in general located at higher distance
 having a median  redshift of z=0.5 (see Fig. \ref{z}). Applying the Wilcoxon,
 Mann, Whitney rank sum test, we find that the probability of BL objects
 having a median redshift larger than that of the NL objects is higher than
 90\%. Again, this is not a highly significant result, but it is highly
 suggestive that NL objects are rare at redshifts higher than $\sim 0.3$.
 Alternatively, as there is a tight correlation  between redshift and
 luminosity in flux limited surveys, high luminosity sources, i.e.  $\rm
 L_x>10^{44}$ \lunits, appear to be    associated primarily with BL AGN while
 the lower  luminosity ones with NL AGN.   The only highly luminous ($\rm
 L_X\sim 10^{45}$ \lunits) NL object in our sample (AXJ0144.9-0345)   has high
 column density ($\rm N_H >10^{22}$ \cunits) and thus is a candidate type II
 QSO.

\begin{table*}
\scriptsize
\begin{center}
\begin{tabular}{ccccccccccc}
\hline
name & $N_{H,Gal}^{1}$ & counts & $\Gamma^2$ & $N_H^3$   & $\chi^2_\nu$ & F(2-10)$^4$ & F(5-10)$^5$ & $L_x^6$ & ID & z \\  
\hline 
AXJ0140.1+0628$^{a,c}$ & 4 & 47  & 1.9 &$12^{+3}_{-5}$	& - & 1.9 & 1.3 &14 & BL &  0.504 \\
AXJ0144.9--0345& 3 & 249 & 1.9 & $2.1^{+0.2}_{-0.4}$ & - & 5.1 & 2.5 & 71 & NL & 0.620 \\
AXJ0335.2--1505& 5 & 124 & 1.9& $2.0^{+0.3}_{-0.5}$	& -   & 4.0 & 2.0 &  1.8 &NL &0.121  \\ 
AXJ0440.0--4534$^{b}$ & 2 & 70 & 1.9 & $1.7^{+0.5}_{-0.3}$ & - & 2.0 & 0.9 & 0.2 & NL & 0.040 \\
AXJ0836.2+5538$^{b,c}$ & 4 & 89 & 1.9& $0.8^{+0.5}_{-0.8}$ & - & 1.0 & 0.45 & 57 &BL & 1.290  \\
AXJ1035.1+3938$^{b,c}$ & 1 & 30  &1.9 &  $7.0^{+3}_{-2.5}$	& -  & 1.5 & 0.68 & 0.7 & NL &  0.107 \\
AXJ1230.8+1433$^{a}$ & 2 & 172 &1.9 & $0.07^{+0.07}_{-0.03}$ & - & 1.2 & 0.54 & 0.3 &BL & 0.115 \\
AXJ1406.1+2233$^{b}$ & 2 & 36 & 1.9 & $2.9^{+1.1}_{-0.9}$ & - & 1.0 & 0.55 & 0.9 &NL& 0.173 \\
AXJ1511.7+0758 & 2 & 407 & $2.12^{+0.28}_{-0.20}$ & $<0.02$		& 52.0/23 & 2.3 & 1.0 & 0.2 &NL & 0.045\\
AXJ1531.8+2414 & 4 & 436 & $2.42^{+0.21}_{-0.18}$ & $<0.02$ & 38.3/23 & 3.1 & 1.40 & 1.0 & BL & 0.096\\
AXJ1531.9+2420 & 4 & 759  & $1.95^{+0.10}_{-0.09}$ & $<0.02$ & 50.7/40 & 6.0 & 2.70 & 111 &BL & 0.631\\
AXJ1532.5+2415$^{a}$ & 4 & 50 & 1.9 & $1.6^{+0.6}_{-0.6}$ & - & 1.5 & 0.75 & - &  - & - \\  
\hline 
\multicolumn{6}{l}{$^1$ Galactic column density in units $10^{20}$ \cunits;} \\
\multicolumn{6}{l}{$^2$ Photon index} \\
\multicolumn{6}{l}{$^3$ Intrinsic column density in units $10^{22}$ \cunits;} \\
\multicolumn{6}{l}{$^4$ Absorbed Flux (2-10 keV) in units $10^{-13}$ \funits;} \\
\multicolumn{6}{l}{$^5$ Absorbed Flux (5-10 keV) in units $10^{-13}$ \funits} \\
\multicolumn{6}{l}{$^6$ Intrinsic Luminosity (2-10 keV) in units $10^{43}$ \lunits} \\
\multicolumn{6}{l}{$^a$ Observed and detected by ROSAT} \\
\multicolumn{6}{l}{$^b$ Observed but not detected by ROSAT} \\
\multicolumn{6}{l}{$^c$ Presented in Nandra et al. 2004} \\
\end{tabular}
\end{center}
\caption{The \chandra spectral fits}
\label{chandrafits} 
\end{table*}

\begin{table*}
\scriptsize
\begin{center}
\begin{tabular}{cccccccc}
\hline
name & $N_{H,Gal}^{1}$ & $\Gamma^2 $ & $N_H^3$   & $\chi^2_\nu$       & F(2-10)$^4 $ & ID & z \\  
\hline 
AXJ0140.1+0628 & 4 & 1.9 & $8.5^{+2.9}_{-2.6}$  & 25.5/22 & 1.3 & BL &  0.504 \\
AXJ0836.2+5538 & 4 & $1.73^{+0.19}_{-0.16}$ & $<0.2$  & 24.1/29 & 0.9 & BL & 1.290  \\
AXJ1230.8+1433 & 2 & 1.9 & $<0.02$ & - & 0.7  & BL & 0.115 \\
AXJ1406.1+2233 & 2 & 1.9 & $3.2^{+1.0}_{-1.0}$ & - & 0.4 & NL& 0.173 \\
AXJ1531.8+2414 & 4 & $2.43^{+0.12}_{-0.08}$ & $<0.02$ & 79.1/90 & 2.8 & BL & 0.096\\
AXJ1531.9+2420 & 4 & $2.14^{+0.07}_{-0.07}$ & $<0.02$ & 148.6/136 & 6.0 & BL & 0.631\\
AXJ1532.5+2415 & 4 & 1.9 & $3.0^{+1.5}_{-1.1}$ & - & 0.5 & - & - \\  
\hline 
\multicolumn{6}{l}{$^1$ Galactic column density in units $10^{20}$ \cunits;} \\
\multicolumn{6}{l}{$^2$ Photon index} \\
\multicolumn{6}{l}{$^3$ Intrinsic column density in units $10^{22}$ \cunits;} \\
\multicolumn{6}{l}{$^4$ Absorbed Flux (2-10 keV) in units $10^{-13}$ \funits;} \\

\end{tabular}
\end{center}
\caption{The \xmm spectral fits}
\label{xmmfits} 
\end{table*}

\begin{figure*}
\rotatebox{270}{\includegraphics[width=4cm]{axj0140.eps}} 
\rotatebox{270}{\includegraphics[width=4cm]{01449_0345.eps}} 
\rotatebox{270}{\includegraphics[width=4cm]{0335.2_1505.eps}} \hfill \\ 
\rotatebox{270}{\includegraphics[width=4cm]{0440.0_4534.eps}} 
\rotatebox{270}{\includegraphics[width=4cm]{axj0836.2.eps}} 
\rotatebox{270}{\includegraphics[width=4cm]{1035.1_3938.eps}} \hfill \\ 
\rotatebox{270}{\includegraphics[width=4cm]{axj1230.8.eps}} 
\rotatebox{270}{\includegraphics[width=4cm]{axj1406.1.eps}} 
\rotatebox{270}{\includegraphics[width=4cm]{1511.7_0758.eps}} \hfill \\
\rotatebox{270}{\includegraphics[width=4cm]{axj1531.8.eps}} 
\rotatebox{270}{\includegraphics[width=4cm]{axj1531.9.eps}} 
\rotatebox{270}{\includegraphics[width=4cm]{axj1532.5.eps}} \hfill \\
 \caption{The X-ray spectra (\xmm or \chandra) together with the best fit power-law 
model and the residuals (lower panels). Only the \xmm spectra are 
 presented (both the MOS and PN in the same panel) where available. 
 In the case where the spectra 
 have been fitted with C-statistic we bin the data to produce 
 a signal to noise ratio of 10 for illustrative purposes 
 }
 \label{xspec}
\end{figure*}

\begin{figure}
\rotatebox{270}{\includegraphics[width=6.5cm]{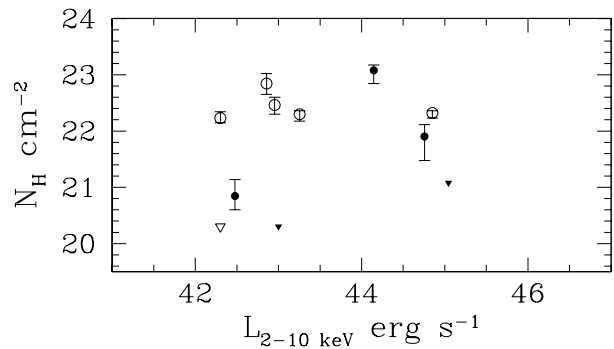}}
\caption{The column density derived from the \chandra spectral fits 
 as a function of 
 the 2-10 keV intrinsic luminosity. BL (NL) AGN are denoted with 
 filled (open) symbols. The triangles denote upper limits. 
 The errors correspond to the 90 per cent confidence level. 
 }
\label{lxnh}
\end{figure}

\section{Discussion}
 
 Our survey provides a glimpse of bright, nearby examples  of the hard AGN
 population which produces an  appreciable fraction of the X-ray background.
 The subsample of the SHEEP survey presented here  contains both Type I (BL)
 and Type II (NL) AGN, according to the optical classification. 

 Although the number of objects that we study is small, we find that there is a
 considerable difference in their redshifts, with NL objects preferentially
 found at redshifts lower than those of BL AGN.  The cross-correlation of {\it
 GINGA} and {\it HEAO-1} X-ray background maps above 2 keV with nearby galaxy 
 catalogues, demonstrates that an appreciable fraction  of the hard X-ray
 emission may arise at low redshift, $z\leq0.1$ (Jahoda et al. 1991, Lahav et
 al. 1993).  Our NL AGN clearly provide examples of this  nearby AGN
 population.    The distance separation between NL and BL AGN  is ubiquitous in
 hard X-ray surveys with the redshift peak  of the populations depending on the
 surveys's flux limit.  

 Our results are consistent with
 the hypothesis that NL AGN are, on average, less luminous than $10^{44}$
 \lunits.  Although our statistics is clearly small, the same  effect i.e. 
 the NL AGN  occupying
 preferentially the low luminosities, can be also witnessed in the \xmm hard
 4.5-7.5 keV sample of Caccianiga et al. (2004)  which contains 28 sources.
 Five out of the seven NL AGN in Caccianiga et al. (2004) have 2-10 keV
 intrinsic luminosities lower than $10^{44}$ \lunits.  Obviously, larger
 surveys are needed in order to confirm unambiguously that  NL AGN are
 intrinsically less luminous than the BL AGN. For example, this could be
 the case in a scenario where highly luminous AGN (QSO) come preferentially in 
 a ``Broad Line flavour" in contrast to the low luminosity  AGN (Seyferts)
 which  appear as both NL or BL according to the viewing angle in respect to
 the obscuring torus. Steffen et al. (2003)
 find that the fraction of NL AGN decreases at high redshift, $z>1$. 
 Treister et al. (2004) observe the same effect but 
 raise the criticism that this could be caused by 
 incompleteness in the spectroscopic identification.  
 Indeed luminous NL AGN at high redshift cannot be easily 
 detected in the optical owing to large amounts of redenning
 (e.g. Fiore et al. 2003). 
 The results emerging from our own survey as well as from  
 that of Caccianiga et al. (2004), which has practically complete spectroscopic 
 identification,  lend support to the findings of Steffen et al. (2003).     
 We note that the density of NL AGN in Steffen et al. (2003) decreases
 only at $z>1$. A direct comparison of the redshift distribution of 
 our sample with that of Steffen et al. 
  is not straightforward as the flux limit of the latter sample is much deeper.    
 Nevertheless, if the NL AGN are intrinsically fainter than the 
 BL AGN, the effect is that the NL AGN are observed at lower redshifts 
 compared with the BL ones in any given survey. 
 Similarly, recent surveys (e.g. Barger et al. 2004) show evidence 
 of ``cosmic downsizing'' where the low luminosity AGN 
 have been formed at a later epoch compared with the 
 bright QSOs. In particular, the peak of the redshift distribution 
 of the low luminosity AGN peaks around z=0.7. 
 Thus given that the NL AGN are low luminositiy sources this provides an 
 additional reason why these are observed at relatively low redshifts.

 The column density distribution is almost equally divided  between high and
 low column densities (table \ref{chandrafits}). In particular the fraction of sources 
 with $N_H>10^{22}$ \cunits is $58\pm22$ per cent. It has to be mentioned that
 we are dealing with a biased  'hard' sample  since we have primarily observed
 with \chandra, the  sources for which no {\it ROSAT} positions are available
 (ie those either not detected or not  observed by {\it ROSAT}) and hence the above 
 fraction should be only viewed as an upper limit.  
 Therefore, our sample cannot provide a fair comparison with the 
 X-ray background population synthesis models 
 of e.g. Comastri et al. (2001). 
 We note that Caccianiga et al. (2004) find a fraction of 
 $26\pm10 $ for the fraction of obscured AGN, inconsistent 
 with the predictions of Comastri et al. (2001).    
We should be able to provide stronger constraints, when
 all the objects in the SHEEP survey have been observed by \chandra.

 We find that NL objects are more obscured compared to BL objects, as
 expected from Unification models.  However, there is a  notable exception as 
 the hardest source  in the sample is associated with a BL AGN (0140.1+0628). 
 The \xmm spectrum presented here confirms  the column density derived from 
 the \chandra hardness ratio analysis (Nandra et al. 2004). 
 The observed colour (B--R=0.8) suggests little absorption  in the optical:
 $A_V\approx1.5$ assuming B--R=0.3 at z=0.5 according to the  SDSS QSO template
 spectrum of Vanden Berk et al. (2002).   On the other hand the column density
 measured  in the X-ray spectrum corresponds  to $A_V\sim 55$ assuming the
 Galactic  dust-to-gas ratio (Bohlin, Savage \& Drake 1978).  This large
 discrepancy can be naturally explained by sublimation  of the dust (eg
 Granato, Danese \& Franceshini 1987).  This may be the case in
 Broad-Absorption-Line (BAL) QSOs  which usually present large absorbing column
 densities  in the X-ray spectra, with typically  $N_H>10^{23}$ \cunits ~
 (Gallagher et al. 2002), while they present moderate reddening in their
 colours (Brotherton  et al. 2001). Actually, from the existing  optical
 spectrum alone we cannot exclude the possibility that our object is a BAL. The
 most efficient diagnostic  for the classification of a source as a BAL is  the
 presence of a broad CIV line (1549 \AA)  which is outside the spectral window
 at the redshift  of  our object. 

 Finally,  one the softest X-ray sources is  associated with a NL source
 (AXJ1511.7+0758). It is puzzling why the  optical reddening which should be 
 responsible for the  obliteration of the broad-line region does not produce 
 significant absorption in X-rays. Typically, NL   AGN (Seyfert-1.8-1.9.-2.0)
 present column densities  higher than $10^{23}$ \cunits (eg Awaki et al.
 1991). One possibility is that at the low luminosity regime   - our source has
 $\rm L_x\sim 10^{42}$ \lunits - no broad-line  region is formed (Nicastro
 2000). Indeed, there are a few   bona-fide examples of Seyfert-2 galaxies, 
 (see Panessa \& Bassani 2002,  Georgantopoulos \& Zezas 2003) with  no
 obscuration in X-rays.  Alternatively, the possibility that, at least, 
 the X-ray emission
 comes  primarily from star-forming processes cannot be ruled out. 
 As discussed earlier the X-ray spectrum requires the presence of a 
 soft thermal component.  This would
 naturally explain the relatively low level  of X-ray luminosity (see e.g.
 Zezas, Georgantopoulos \& Ward 1998)  as well as the soft X-ray spectrum.   
 Nevertheless, the weak $H\beta$ relative to the [OIII] line as well 
 as the relative strength of the NII relative to the Ha suggest against 
 the star-formation activity.

 \section{Summary} 
 We present \chandra and \xmm observations of a subsample 
 of 12 bright sources ($\rm f(2-10 keV) >10^{-13}$ \funits)
  from the SHEEP \asca 5-10 keV catalogue. We have obtained 
 optical spectroscopic observations for 11 of these.
 The optical spectra show that our sources are associated 
 with both NL and BL AGN. The X-ray spectra reveal intrinsic absorbing 
 column densities of the order $10^{20}-10^{23}$ \cunits.
 The BL AGN on average  
 do not present obscuration in X-rays compared to the 
 NL AGN. There are however exceptions in this trend with the
 most obscured source being a BL AGN.
 Although the statistics is limited,     
 the NL AGN are found at lower redshift 
 and luminosity compared with the BL sources.
 The most likely explanation is that NL AGN are intrinsically 
 less luminous than the BL ones and thus they 
 are preferentially found  at lower redshifts.

\section{Acknowledgments}
 
 This work is funded by the Greek National Secretariat for Research 
 and Technology in the framework of the Programme Participation  
 in Projects of International Organisations (European Space Agency). 
 We also acknowledge support by the European
 Union and the Greek Ministry of Development  in the framework of the
 programme  'Promotion of Excellence in Technological Development  and
 Research', project 'X-ray Astrophysics with ESA's mission XMM'. We
 acknowledge the  use of data from the {\it XMM-Newton} Science Archive at
 VILSPA. Skinakas Observatory is a collaborative project of the University of
 Crete, the Foundation for research and Technology-Hellas, and the
 Max-Planck-Institut f\"{u}r Extraterrestrische Physik.

\end{document}